\documentclass[11pt]{article}
\pdfoutput=1
\usepackage{amsmath,amssymb,graphicx}
\usepackage{hyperref}
\usepackage{palatino}
\usepackage{mathpple}
\usepackage[font=small,labelfont=bf]{caption}
\usepackage{cite}
\usepackage{multirow}
\usepackage{slashed}

\newcommand{\beq}{\begin{eqnarray}}
\newcommand{\eeq}{\end{eqnarray}}

\usepackage[usenames,dvipsnames]{xcolor}

\usepackage{tikz}
\usetikzlibrary{trees}
\usetikzlibrary{decorations.pathmorphing}
\usetikzlibrary{decorations.pathreplacing}
\usetikzlibrary{decorations.markings}
\usetikzlibrary{positioning,arrows,patterns}
\usetikzlibrary{calc}
\usetikzlibrary{shapes.misc}
\tikzset{
    photon/.style={decorate, decoration={snake}, draw=black},
    electron/.style={draw=black, postaction={decorate},
        decoration={markings,mark=at position .55 with {\arrow[draw=black]{>}}}},
    majfermion/.style={draw=black},        
    scalar/.style={draw=black,dashed},
    directedscalar/.style={draw=black, dashed,postaction={decorate},
        decoration={markings,mark=at position .55 with {\arrow[draw=black]{>}}}},
    gluon/.style={decorate, draw=black,
        decoration={coil,amplitude=4pt, segment length=5pt}},
    majorana/.style={draw=black},
    cross/.style={cross out, draw=black, minimum size=1pt, inner sep=2pt, outer sep=2pt},
    cross/.default={1pt}    
}

\newcommand{\semiloop}[4][]{%
        \draw[#1] let \p1 = ($(#3)-(#2)$) in (#3) arc (#4:({#4+180}):({0.5*veclen(\x1,\y1)});) 
}

\newcommand{\semiloopback}[4][]{%
        \draw[#1] let \p1 = ($(#3)-(#2)$) in (#3) arc (#4:({#4-180}):({0.5*veclen(\x1,\y1)});) 
}

\title{\bf Nonthermal production of dark radiation and dark matter}
\author{Matthew Reece and Thomas Roxlo\\
{\small \color{gray} \texttt{mreece, troxlo@physics.harvard.edu}}\\
{\em Department of Physics, Harvard University, Cambridge, MA 02138, USA}}

\begin{document}
\maketitle

\begin{abstract}
Dark matter may be coupled to dark radiation: light degrees of freedom that mediate forces between dark sector particles. Cosmological constraints favor dark radiation that is colder than Standard Model radiation. In models with fixed couplings between dark matter and the Standard Model, these constraints can be difficult to satisfy if thermal equilibrium is assumed in the early universe. We construct a model of asymmetric reheating of the visible and dark sectors from late decays of a long-lived particle (for instance, a modulus). We show, as a proof of principle, that such a model can populate a sufficiently cold dark sector while also generating baryon and dark matter asymmetries through the out-of-equilibrium decay. We frame much of our discussion in terms of the scenario of dissipative dark matter, as in the Double-Disk Dark Matter scenario. However, our results may also be of interest for other scenarios like the Twin Higgs model that are in danger of overproducing dark radiation due to nonnegligible dark-visible couplings.
\end{abstract}

\section{Introduction}

The Standard Model involves multiple forces and many particles, but accounts for a little less than one-sixth of the matter content of the universe~\cite{Ade:2013sjv}. Theorists often postulate that the remaining matter consists of just one type of cold, noninteracting dark matter particle. Recently a great deal of interest has arisen in the possibility that dark matter instead is a component of a richer ``dark sector'' that could be of comparable complexity to the Standard
Model. In particular, dark matter may couple to massless (or nearly massless) particles that would constitute dark radiation. Constraints from Big Bang Nucleosynthesis~\cite{Cyburt:2004yc,Cyburt:2015mya} and the CMB~\cite{Ade:2013zuv} limit the amount of dark radiation to less than about one extra effective neutrino species. However, the constraints are somewhat stronger if dark radiation interacts strongly with dark matter, due to the induced dark acoustic oscillations~\cite{Cyr-Racine:2013fsa}. As a result, if a dark sector involves new massless degrees of freedom, they must be colder than the visible sector. The temperatures of the dark and visible sectors can be closely related if the two sectors were in thermal equilibrium at some point in the past~\cite{Feng:2008mu,Franca:2013zxa,Brust:2013ova}. This suggests a bleak outlook for direct or indirect detection of any dark matter candidate with significant couplings to dark radiation. Our goal in this paper is to explore to what extent that conclusion is true. In particular, significant departures from thermal equilibrium can apparently weaken the argument, so we will explore nonthermal cosmological production of both the dark and visible sector matter and radiation.

A set of closely related models of dark matter coupled to massless dark photons have been studied in the guises of mirror matter~\cite{Khlopov:1989fj,Berezhiani:1995am,Mohapatra:2000qx,Foot:2004pa,Berezhiani:2005ek,Higaki:2013vuv,Foot:2014mia}, atomic dark matter~\cite{Goldberg:1986nk,Kaplan:2009de,Kaplan:2011yj,Cline:2012is,CyrRacine:2012fz,Cline:2013zca,Cline:2013pca,Cyr-Racine:2013fsa,Buckley:2014hja,Pearce:2015zca}, and a plasma of particles with dark charge~\cite{Feng:2008mu,Ackerman:mha,Feng:2009mn,Petraki:2014uza,Foot:2014uba,Foot:2014osa}. Due to its ability to radiate dark photons which carry energy out of a galaxy, dark matter in these theories is dissipative and can cool and form collapsed structures. In the mirror matter context it has generally been assumed that some heating process restores energy to the dark matter so that it can remain in an extended halo. In other models, parameters have been chosen to avoid significant dissipation. Recently, it was observed that if there are multiple components of dark matter, a small fraction could have dissipative dynamics while evading current constraints~\cite{Fan:2013yva,Fan:2013tia}. This subdominant dissipative dark matter could form galactic disks or other dynamical structures. We refer to this type of dark matter as DDDM, for ``Double Disk Dark Matter.''

We can organize the possible observational consequences of dissipative dark matter into two broad categories: those that depend only on its gravitational influence on baryons, and those that postulate more direct interactions between Standard Model particles and dark sector particles. The former case we will refer to as ``gravitational signatures'' and the latter as ``particle physics signatures.'' Particle physics signatures fall into the familiar categories of direct detection, indirect detection, and production of dark matter at colliders. The gravitational signatures are diverse. Dissipative dark matter that organizes itself into a disk could be detected by the added gravitational force on stars, which enhances the oscillations of stars in the Milky Way disk around the midplane of the galaxy. Such effects were originally pointed out for disks formed by accretion of cold dark matter onto the baryonic disk~\cite{Read:2008fh} and are probed by stellar kinematics measured in surveys like Hipparcos~\cite{Holmberg:2004fj}, SEGUE~\cite{Zhang:2012rsb}, or future data from Gaia.  A dark disk formed by dissipation in the dark sector is potentially thinner than a dark disk formed by accretion~\cite{Fan:2013yva}, which motivates new fits to stellar kinematics data and to data on the distribution of atomic and molecular gas near the galactic plane~\cite{Kramer:2016dqu,McKee:2015hwa,Kramer:2016dew}. The gravitational effect of the disk could influence the loss rate of comets from the Oort cloud~\cite{Randall:2014lxa}. Dissipative interactions could influence galactic structure~\cite{Foot:2013vna,Foot:2014uba}, cluster mergers \cite{Heikinheimo:2015kra}, or even play a role in explaining the unusual spatial geometry of Andromeda's satellite galaxies~\cite{Foot:2013nea,Randall:2014kta}. Another very interesting gravitational signature is the possible effect of a dark-sector accretion disk and corresponding dark-sector jets on the evolution of a black hole's rotational energy, which might be detected through observations of long-duration gamma ray bursts~\cite{Banks:2014rsa,Fischler:2014jda,Fischler:2015zka}. Dark acoustic oscillations and their effect on structure formation are another purely gravitational effect of the dark sector's dynamics on observable properties of the visible sector~\cite{Cyr-Racine:2013fsa,Buckley:2014hja}. These effects are very interesting since they allow us to use the weakest force we know (gravity) as a probe of particle physics interactions in the dark sector that we have no direct access to. This is not a new lesson, being familiar from the case of how self-interacting DM can alter structure on the scales of dwarf galaxies that is then detected through stellar kinematics~\cite{Spergel:1999mh,Rocha:2012jg,Peter:2012jh,Tulin:2013teo}, but it has interesting new aspects in the case with dissipation. If we assume that the dark and visible sectors are coupled only through gravitational-strength interactions, there are interesting scenarios for explaining the relative abundances in the two sectors. An example is \cite{Fischler:2014jda}, which discusses the Affleck-Dine mechanism to generate asymmetries in both sectors and thermal freezeout (with separate visible and dark plasmas at different temperatures) to generate symmetric components.

Particle physics signatures of dissipative dark matter require larger couplings that potentially link the dark radiation energy density in the early universe to that of visible radiation. Detectability of these signatures is thus threatened by bounds from the CMB. These signatures are similar to those of ordinary cold dark matter, but can have interesting new features due to the different location and kinematics of disk dark matter compared to ordinary dark matter. For direct detection, the unusually low velocity of dark matter in a co-rotating disk (relative to the Earth) means that little energy goes into nuclear recoils and signals could easily lie below the threshold of current experiments~\cite{Bruch:2008rx,Fan:2013yva,Fan:2013bea}.  Even large cross sections of order those expected for $Z$ exchange with WIMPs could have evaded detection due to the unusual kinematics. An extended model of dissipative dark matter known as ExoDDDM could even be useful in reconciling existing hints of signals with exclusion bounds~\cite{McCullough:2013jma}. The low velocity of disk dark matter enhances its capture rate in the Sun, leading to interesting prospects for signals of neutrinos from annihilating dark matter inside the Sun~\cite{Bruch:2009rp,Fan:2013bea}. Given how interesting these signals can be, it is important to ask whether they are already indirectly excluded by bounds on dark radiation and dark acoustic oscillations.

The approach that we take builds on several existing ingredients in the literature. Our models are a special case of Asymmetric Dark Matter \cite{Nussinov:1985xr,Kaplan:1991ah,Zurek:2013wia}. The idea of asymmetric reheating, i.e.~that reheating can occur at different temperatures in different sectors, is rather old~\cite{Berezhiani:1995am}, although it is typically discussed for the case of two sectors that interact very weakly. If two sectors interact extremely weakly (e.g.~only with gravitational strength), it is always possible to reheat one while simply diluting any energy stored in the other (e.g.~\cite{Randall:2015xza}). The challenge in our case is that we would like to consider sectors with larger couplings to each other, so any reheating is expected to communicate with both, and furthermore the reheating temperature must be quite low to prevent the two sectors from interacting and thermalizing. The idea of a nonthermal cosmology with decays reheating the universe to a temperature below conventional thermal freezeout temperatures and above the temperature of BBN has a long history, especially in the context of supersymmetry, where particles like winos that annihilate very efficiently make poor thermal dark matter candidates but can be ideal nonthermal dark matter candidates \cite{Moroi:1999zb,Gelmini:2006pw,Acharya:2009zt,Arcadi:2011ev,Kane:2011ih,Moroi:2013sla}. Recent strong indirect detection constraints on conventional SUSY nonthermal dark matter \cite{Cohen:2013ama,Fan:2013faa,Ackermann:2015zua} have sparked new interest in supersymmetric scenarios with hidden-sector dark matter particles \cite{Blinov:2014nla,Kane:2015qea}. If the hidden sector is rich enough, the scenario we discuss in this paper could naturally find a home in such scenarios.\footnote{Non-thermal dark matter could also originate directly in decays of the inflaton, rather than other long-lived cosmological moduli \cite{Dev:2013yza,Dev:2014tla}.}

A model containing both a neutral DM particle and DDDM-like particles charged under a long-range force recently appeared in ref.~\cite{Altmannshofer:2014vra} as part of a model motivated by a radiative origin of the weak scale. The authors observed that the direct detection cross section of their DDDM-like relic particles was small enough as to be lost to the neutrino background. This is a specific example of the general tension that we highlight in section~\ref{sec:tension}.

In section \ref{sec:tension}, we will give a quantitative explanation of the tension between having detectable particle physics signatures of dissipative dark matter and evading cosmological constraints on the coupling of dark matter and dark radiation. Specifically, we will argue that a direct detection cross section of around the LUX bound translates into a large enough coupling to maintain thermal equilibrium between the dark and visible sectors down to a temperature of about 1 GeV. If the two sectors decouple at such late times, there is inevitably too much dark radiation. In section \ref{sec:modulidecay}, we explain how this tension can be ameliorated by a departure from thermal equilibrium. Specifically, we consider the decay of a long-lived particle like a modulus. If its decays reheat the visible sector to a temperature below 1 GeV but above the temperature of BBN, the conclusion of section \ref{sec:tension} may be avoided. Moduli decays also open an interesting opportunity to explain the abundances of both baryons and of asymmetric dark matter via the ``cladogenesis'' mechanism \cite{allahverdi2,allahverdi1}. We examine the prospects for explaining both the baryon asymmetry and the dissipative dark matter asymmetry through this mechanism, while simultaneously maintaining a sufficiently low dark-sector temperature. We find that it is possible. (We also discuss constraints on the theory from absence of observed low-energy baryon number violation.) In section \ref{section:smallnumbers} we offer a few additional comments on the model-building, including how bounds on kinetic mixing of the dark and visible sectors can be avoided. Finally in section \ref{sec:conclusions} we offer concluding remarks.

\section{Tension between CMB constraints and detectability}
\label{sec:tension}

\subsection{Origin of the tension}

Cosmological bounds on DDDM depend strongly on the ratio 
\begin{equation}
\xi \equiv \frac{T_{D}}{T_{\rm SM}}
\end{equation} 
of the temperature of dark radiation to the temperature of the Standard Model photons at the time of last scattering. Cyr-Racine et al.~\cite{Cyr-Racine:2013fsa} carried out a detailed study of cosmological constraints on $\xi$ in terms of a number of factors such as the strength of the dark matter-dark radiation interaction and the fraction of dark matter which interacts with dark radiation. They found that, to reasonable approximation, the bounds on $\xi$ fall into one of two regimes. If
the dark matter-dark radiation coupling is very low, then the constraint on the dark radiation is simply the standard bound on the number of ``effective neutrino species,'' which requires $\xi \lesssim 0.5$ at around 68\% confidence ($\xi \lesssim 0.6$ at 95\% confidence). However, if the coupling rises above a critical threshold, other constraints come into play from the cosmological imprint of the dark radiation upon dark matter through dark acoustic oscillations. In this case, the bound becomes much
stricter: $\xi \lesssim 0.1$ at 68\% confidence ($\xi \lesssim 0.2$ at 95\% confidence). The transition between these two regimes can be quite sharp, as seen, for example, in Figure 12 of \cite{Cyr-Racine:2013fsa}. In this paper we will be concerned with the interacting regime in which the bound is stronger. In particular, we will take $\xi \approx 0.1$ as a target for which there is no tension with cosmological data.

We do not assume that dark matter has dissipation substantial enough to form a disk, merely that it couples to dark radiation strongly enough to affect the cosmological constraint. On the other hand, if dark matter is to form a disk or have other interesting structure due to its interaction with dark radiation, it will almost certainly fall into the strongly-interacting regime. However, the ratio of temperatures $\xi$ is closely related to when DDDM decoupled from ordinary matter
in a thermal cosmology. This, in turn, relates to how tightly the two sectors are coupled, which has an impact on possible signals from direct detection or solar capture~\cite{McCullough:2013jma, Fan:2013bea}. Our goal in this section is to look at various models and understand to what extent the cosmological constraints preclude the possibility of detecting DDDM through its non-gravitational interactions with normal matter. We will see that it is difficult or impossible for a normal thermal history to simultaneously result in a small $\xi$ and a detectable interaction cross-section. In subsequent sections, we will show that this can be achieved in a non-thermal decay scenario.

If the DDDM follows a normal thermal cosmology and was at some point in thermal contact with ordinary matter, then the simplest way to lower its temperature is to add particles to the visible sector which have fallen out of equilibrium since the dark matter decoupled. Since the entropy density, $s \sim g_{*} T^3$, is separately conserved in the different sectors after they decouple, we have \cite{Fan:2013yva}
\begin{equation}
\frac{g_{*s,D}^{\text{dec}}}{g_{*s,D}(t)\xi(t)^3} = \frac{g_{*s,\text{vis}}^{\text{dec}}}{g_{*s,\text{vis}}(t)}
\end{equation}
\begin{equation}
\xi(t) = \left(\frac{g_{*s,D}^{\text{dec}}}{g_{*s,D}(t)} \frac{g_{*s,\text{vis}}(t)}{g_{*s,\text{vis}}^{\text{dec}}}\right)^{1/3} \lesssim 0.1
\label{eq:xi}
\end{equation}
Here subscript $D$ refers to the dark sector and superscript ``dec'' refers to quantities at the time that the visible and dark sectors decouple from each other.

Assuming that neither sector undergoes recoupling, so that $g_{*s}^{\text{dec}} > g_{*s}(t)$, the only way to achieve such a low $\xi$ is to increase $g_{*s,\text{vis}}^{\text{dec}}$. We assume that the dark sector at low temperatures has only the fermions $\chi$ and ${\overline \chi}$ (which may be thought of as a ``dark electron'' and ``dark positron'') and the U(1) dark photon $\phi$, and that the dark fermions froze out from the dark radiation some time since the two sectors decoupled, ie. $g_{*s,D}^{\text{dec}} = 2 + \frac{7}{8} \times 4 = 5.5$, $g_{*s,D}(t) = 2$. Then we must have
\begin{equation}
\frac{5.5}{2} \frac{g_{*s,\text{vis}}(t)}{g_{*s,\text{vis}}^{\text{dec}}} \lesssim \frac{1}{1000}
\end{equation}
or, given 2 present day visible degrees of freedom from the photon, $g_{*s,\text{vis}}^{\text{dec}} \gtrsim 5500$. This is not very likely. However, if we allow $\xi \lesssim 0.25$ instead, which may just barely evade the bounds on dark acoustic oscillations \cite{Cyr-Racine:2013fsa}, we ``only'' need roughly 352 degrees of freedom in the visible sector at decoupling. This may, very optimistically, be possible due to beyond the Standard Model physics at temperatures
slightly above the current LHC bounds, say $T_{\text{dec}} = 10 \text{ TeV}$.

Now we come to the basic origin of the tension: decreasing $\xi$ requires that many Standard Model degrees of freedom fall out of thermal equilibrium with the Standard Model {\em after} the dark radiation does. However, the tighter the coupling between dark sector particles and the Standard Model, the longer they will stay in equilibrium with one another. Thus the CMB constraints prefer small couplings and an early decoupling, whereas a detectable signal in direct detection experiments prefers late decoupling.

\subsection{Quantifying the tension}

To estimate the relationship between the decoupling temperature and the direct detection cross-section, we'll begin with a toy model of dark matter interacting with normal matter through higher-dimension operators. The dimension-six relativistic operators that lead to direct detection rates that are not momentum suppressed are ${\bar \chi}\chi {\bar q}q$ or ${\bar \chi}\gamma^\mu \chi {\bar q}\gamma_\mu q$; further operators that give smaller direct detection rates are enumerated in Refs.~\cite{Fan:2010gt,Fitzpatrick:2012ix,Fitzpatrick:2012ib}. Thus, let us consider an effective interaction
\beq
{\cal L}_{\rm eff} = \sum_{q} \left(\frac{1}{\Lambda_{0;q}^2} {\bar \chi}\chi {\bar q}q + \frac{1}{\Lambda_{1;q}^2} {\bar \chi}\gamma^\mu \chi {\bar q}\gamma_\mu q\right).
\label{eq:Leff}
\eeq
Depending on the flavor structure of the couplings and the details of the UV completion, there may be various constraints on the scales $\Lambda_{0;q}$ and $\Lambda_{1;q}$. 

We could also consider models with light, weakly interacting mediators, so that dark matter--baryon interactions are not accurately approximated by contact interactions. However, in that case there are additional processes in the thermal plasma that can equilibrate dark matter and baryons via their mutual interactions with mediators in the plasma, e.g. ${\overline \chi} \chi \leftrightarrow \phi~({\rm or}~\phi\phi)$ and $q {\overline q} \leftrightarrow \phi~({\rm or}~\phi\phi)$. Hence, we expect that decoupling will tend to be even later in the presence of such light mediators. Furthermore, the mediators themselves potentially impose additional cosmological constraints. Hence, although we will not address this scenario in great detail, we expect that the tension we will derive in the effective contact interaction scenario should persist in the case of light mediators. One might wonder about a possible caveat when direct detection today is enhanced by inverse powers of velocity, whereas the thermal medium might have existed when the average velocity was larger. But this requires very light mediators that are relativistic at late times and hence is strongly constrained cosmologically as well as by low-energy experiments, so it seems unlikely to offer an easy way to evade the tension.

For simplicity, let us first focus on a coupling to one particular quark flavor, $\Lambda \equiv \Lambda_{0;q}$, with the other couplings negligible. The general case may be obtained by replacing $\Lambda$ in the equations below with an effective value that sums the contributions of all the operators in eq.~\ref{eq:Leff} to the total scattering rate $\Gamma$. There are two regimes in which decoupling can happen: either it occurs when the dark fermions are still relativistic (in which case it is closely analogous to the decoupling of neutrinos in the Standard Model) or it occurs when the dark matter fermions are nonrelativistic and their abundance is Boltzmann-suppressed (analogous to standard WIMPs). Let us consider the former case first. In the early universe, at high temperatures $T \gg m_\chi, m_q$, we can calculate that the rate for dark matter--quark scattering goes as
\beq
\Gamma = n\left<\sigma v\right> = \frac{\zeta(3)}{\pi^3} \frac{T^5}{\Lambda^4} \approx 0.039 \frac{T^5}{\Lambda^4},
\eeq
where $\zeta(3) \approx 1.2$ is the Riemann zeta function of 3. We compare this to the Hubble expansion rate
\beq
H = \sqrt{\frac{\pi^2 g_*}{90}} \frac{T^2}{M_p}.
\eeq
Decoupling happened when $H \approx \Gamma$, i.e.~at the temperature
\beq
T_{\rm dec} \approx \left(\sqrt{\frac{\pi^2 g_*}{90}} \frac{\Lambda^4}{M_p} \frac{\pi^3}{\zeta(3)}\right)^{1/3} \approx 19~{\rm GeV} \left(\frac{\Lambda}{100~{\rm TeV}}\right)^{4/3} \left(\frac{g_*}{352}\right)^{1/6}.
\label{eq:decouplinglight}
\eeq
or, conversely, that to obtain a desired decoupling temperature we require the coupling to be suppressed by
\begin{equation}
\label{2:lambdadef}
\Lambda \approx 1.1 \times 10^7 \text{ GeV} \left(\frac{T_{\text{dec}}}{10 \text{ TeV}} \right)^{3/4} \left(\frac{352}{g_*}\right)^{1/8}.
\end{equation}
Again, this calculation is approximately correct if the abundance of the $\chi$ particles is not Boltzmann-suppressed at the time of decoupling: $m_\chi \lesssim T_{\rm dec}$. For the dark electron masses considered in refs.~\cite{Fan:2013yva,Fan:2013tia}, this is a reasonable assumption. For the heavier dark proton, it may not be. If we consider a model in which the dark electron has negligible scattering cross section but the dark proton scatters, we find that the dark sector decouples at a temperature of
\begin{equation}
T_{\rm dec} \approx \frac{m_\chi}{29 - 4 \log(\Lambda/{\rm TeV})}.
\label{eq:decouplingheavy}
\end{equation}
Thus, for $\Lambda \sim {\rm TeV}$, if we would like the dark proton to decouple at temperatures above 1 TeV we would need its mass to be at least about 30 TeV. Already when the dark proton mass is 100 GeV there is limited parameter space for cooling, and decoupling happens near the GeV scale where $g_{*s}$ is known and too small to lead to small $\xi$. Furthermore, for a fixed fraction of the total dark matter abundance, larger dark matter masses imply smaller number densities and hence lower rates of scattering in direct detection experiments. For all of these reasons, the regime in which eq.~\ref{eq:decouplingheavy} applies is of little interest for us in this paper.

It could also be the case that dark matter scatters primarily with a heavy Standard Model particle that decouples: for instance, we could have a coupling dominantly to the top quark. In this case, a formula like Eq.~\ref{eq:decouplingheavy} might apply but with the top quark mass appearing in the numerator. However, this leads to small enough $T_{\rm dec}$ that we are stuck with large $\xi$. For the purposes of our argument, it suffices to always take the decoupling temperature to be the larger of equations~\ref{eq:decouplinglight} and~\ref{eq:decouplingheavy}, rather than solving the Boltzmann equation for a more precise interpolation between the two.

Now, what does the scale $\Lambda$ mean for dark matter--nucleon scattering at a direct detection experiment? Bounds are usually set on a DM--nucleon scattering cross section
\beq
\label{2:sigmadef}
\sigma = \frac{4 \mu_n^2}{\pi} \left|f_n\right|^2,
\eeq
where $\mu_n = m_\chi m_n/(m_\chi + m_n)$ is the reduced mass of the DM--nucleon system, and $f_n$ is an effective coupling of the DM to the nucleon. This coupling depends on nucleon matrix elements that are used to relate couplings to quarks to couplings to nucleons, as reviewed in~\cite{Belanger:2008sj,Belanger:2008gy}. For the scalar matrix elements of quarks in a nucleon (neutron or proton),
\beq
f^n_{0;q} \equiv \left<n \middle| m_q {\bar q} q \middle| n\right>/m_n,
\eeq 
we will use the lattice-based results $f^n_{0; u} = f^n_{0;d} = 0.025$ and $f^n_{0; s} = 0.053$~\cite{Giedt:2009mr,Cheung:2012qy}. The vector matrix elements are simply determined by current conservation and count the valence up and down quarks in the proton or neutron. (Heavy quarks coupling through vector currents do contribute to direct detection, but only through a dimension-seven gluonic operator whose effect is relatively small~\cite{Fan:2010gt,Kaplan:1988ku}.) The DM--nucleon coupling, summing over all contributions, is given by:
\beq
f_n & = & \frac{1}{2} \left(\sum_{q=u,d,s} \frac{f^n_{0;q}}{\Lambda^2_{0;q}} \frac{m_n}{m_q} + \frac{2}{27} f^n_{0;G} \sum_{q = c,b,t} \frac{1}{\Lambda^2_{0;q}} \frac{m_n}{m_q} \pm \frac{n_u}{\Lambda_{1;u}^2} \pm \frac{n_d}{\Lambda_{1;d}^2}\right), 
\eeq
where $f^n_{0;G} = 1 - \sum_{q=u,d,s} f^n_{T_q}$ and $n_u = 2, n_d = 1$ for the proton and $n_u = 1, n_d = 2$ for the neutron. The overall factor of $1/2$ arises because our dark matter is a Dirac fermion~\cite{Belanger:2008sj} and the sign $\pm$ depends on whether the particle scattering is the dark fermion or its antiparticle, since the vector current ${\bar \chi} \gamma^\mu \chi$ has opposite sign in the two cases.

Let us give numerical values for $f_n$ for a few simplified scenarios. We are not giving detailed models or examining other constraints on these scenarios, but they should be representative of the range of outcomes:
\begin{itemize}
\item Suppose DM has scalar couplings only to second generation (strange and charm) quarks with equal strength, $\Lambda_{0;s} = \Lambda_{0;c} = \Lambda$. Then we compute 
\beq
f_n = \frac{1}{2\Lambda^2} \left(0.0532 \frac{m_n}{m_s} + \frac{2}{27} \left(1 - 0.025 - 0.025 - 0.0532\right) \frac{m_n}{m_c}\right) \approx \frac{0.29}{\Lambda^2}.
\eeq
In fact, in this case the strange quark contribution dominates over charm, contributing $f_n \approx 0.26/\Lambda^2$.
\item If DM scatters with quarks through Higgs exchange, we can take $\frac{1}{\Lambda_{0;q}^2} = \frac{y_\chi y_q}{m_h^2}$, where $y_q = \sqrt{2} m_q/v$ is the quark's Yukawa coupling and $y_\chi$ is the coupling of dark matter to the Higgs. In this case we find that
\beq
f_n \approx 8 \times 10^{-4} \frac{y_\chi}{m_h^2}. 
\eeq
In this case the effective $\Lambda$ in the early universe will be very sensitive to the decoupling temperature; for instance, if $T_{\rm dec}$ is greater than the top quark mass then $\Lambda_{\rm eff} \sim m_h/\sqrt{y_\chi}$, but at lower temperatures only quarks with small Yukawa couplings are available to scatter with and the effective $\Lambda$ becomes larger.
\item If DM scatters through a $Z'$ that couples to both dark matter and baryon number, we can take $\Lambda_{1;d} = \Lambda_{1;u} = \Lambda$ and
\beq
f_n = \frac{3}{2 \Lambda^2}.
\eeq
\end{itemize}

Based on these examples, we can see that if $\Lambda_{\rm eff}$ is the relevant scale for decoupling of the dark and visible sectors in the early universe---the scale appearing in eqs.~\ref{eq:decouplinglight} and~\ref{eq:decouplingheavy}---then the direct detection cross section will tend to be {\em at most} of order $\mu_n^2 / \Lambda_{\rm eff}^4$. In other words,
\begin{equation}
\label{2:sigmadef}
\sigma \lesssim \frac{\mu_n^2}{\Lambda_{\rm eff}^4} \approx 4 \times 10^{-44}~{\rm cm}^2~\left(\frac{10~{\rm TeV}}{\Lambda_{\rm eff}}\right)^4.
\end{equation}
This shows that for $\Lambda_{\rm eff} \sim 10~{\rm TeV}$, the cross section could be in the range currently being probed: the LUX constraint reaches down to $7.6 \times 10^{-46}~{\rm cm}^2$ for dark matter masses of about 30 GeV~\cite{Akerib:2013tjd,Szydagis:2014xog}. However, to achieve a large enough $g_*$ at decoupling (and hence small enough $\xi$ to satisfy CMB constraints) we expect that we will need at least $T_{\rm dec} \gtrsim 1~{\rm TeV}$ and hence, from eq.~\ref{2:lambdadef}, $\Lambda_{\rm eff} \gtrsim 1000~{\rm TeV}$, leading to a direct detection cross section several orders of magnitude too small to be detected in any near-future experiment. Any detection would have to contend with the neutrino floor.

\begin{figure}[h]
\begin{center}
\includegraphics[width=0.75\textwidth]{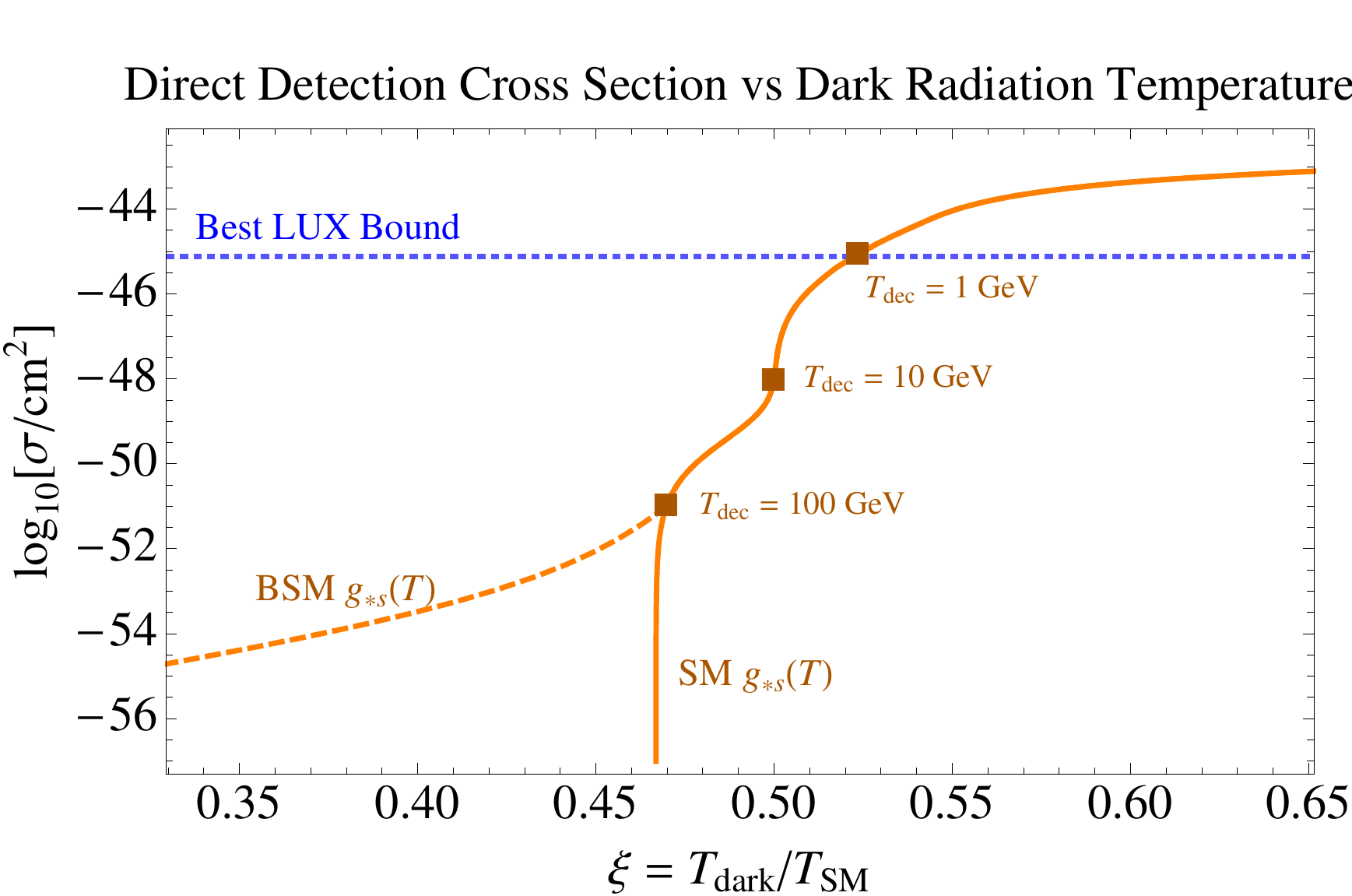}
\end{center}
\caption{The tension between the cosmological preference for a small dark radiation temperature ratio $\xi$ and a detectable dark matter--nucleon direct detection cross section $\sigma$. Although physics beyond the Standard Model can modify the count of degrees of freedom and bend $\xi$ toward smaller values, as shown by the dashed orange BSM curve, this can only happen at temperatures above about 100 GeV, which already correspond to undetectably small cross sections.}
\label{fig:tension}
\end{figure}

To make the tension clearer, we combine the equations \ref{2:lambdadef} and \ref{2:sigmadef} to obtain
\beq
\sigma \lesssim \frac{\mu_n^2}{\Lambda_{\rm eff}^4} \approx 2.3 \times 10^{-56} \text{ cm}^2 \left(\frac{10 \text{ TeV}}{T_{\text{dec}}}\right)^3 \left(\frac{g_*(T_{\rm dec})}{352}\right)^{1/2}.
\eeq
For a given decoupling temperature, we can compare this cross section $\sigma$ with the dark radiation temperature ratio $\xi$ from eq.~\ref{eq:xi}. We use a tabulation of $g_*(T)$ and $g_{*s}(T)$ in the Standard Model from the DarkSUSY code~\cite{Gondolo:2004sc}, which is particularly useful for the regime around the QCD scale. The result is shown in Figure~\ref{fig:tension}. The solid orange curve shows the attainable $\xi$ including only Standard Model degrees of freedom. The dashed orange curve allows for new physics beyond the Standard Model (BSM) to add to $g_{*s}(T)$. We have chosen to add a new degree of freedom for every 10 GeV energy increase above 100 GeV:
\beq
g^{\rm BSM}_{*}(T) = \begin{cases} 
                    g^{\rm SM}_{*}(T) & \text{if}\ T \leq 100~{\rm GeV} \\
                    g^{\rm SM}_{*}(T) + \frac{T-100~{\rm GeV}}{10~{\rm GeV}} & \text{if}\ T > 100~{\rm GeV.}
                                         \end{cases}
\eeq
This is a rapid growth in the number of states, but we see from Fig.~\ref{fig:tension} that it is hopeless: already, when the decoupling temperature is 100 GeV, the largest attainable cross section is around 10$^{-50}$ cm$^2$. To reach smaller values of $\xi$ at cross sections that are near the current bounds, we would have to significantly alter the density of states already between 1 and 10 GeV. Given that collider experiments have thoroughly explored this energy regime, this does not seem possible.

The LUX constraint on the dark matter--nucleon scattering cross section assumes a standard Maxwellian velocity distribution and a dark matter density of $\rho \approx 0.3~{\rm GeV}/{\rm cm}^3$ in the vicinity of the Earth. In the case of dissipative dark matter, the disk of dark matter could have significantly higher density near the galactic plane, and would also scatter with lower velocity. This will change the precise bound. However, these effects are not expected to be dramatic enough to resolve the several orders of magnitude of tension that we have found.

Figure~\ref{fig:tension} makes it clear that, in order to have direct detection cross sections in the regime currently being probed by experiments, the universe must have been out of equilibrium at temperatures below or of order a GeV. Considering such a scenario will be our next step.

\section{Observable DM with low relative temperature from moduli decay}
\label{sec:modulidecay}

Having ruled out thermal production of appreciable quantities of observable DDDM, we now ask whether the DDDM can be generated from the out-of-equilibrium decay of relic particles. Models of this kind are of course extremely common in the visible sector as explanations for the baryon asymmetry of universe \cite{davidson, covi}. Such a theory is attractive to us both because the lack of thermal equilibrium means that the temperatures of the different sectors need not be
directly linked at the time of reheating and because decaying particles can
efficiently generate a dark matter-antimatter asymmetry. This is useful because a highly asymmetric dark sector, in analogy with baryonic matter, allows us to up the interaction cross-section without worrying about losing the DDDM population to self-annihilation.

Consider a long-lived particle $\tau$, perhaps a string modulus or a sterile neutrino, which at some time when the energy density is much larger than $\left(1~{\rm GeV}\right)^4$ comes to dominate the energy (and, for convenience, entropy) density of the universe. (We emphasize that the notation $\tau$ should not be confused with the tau lepton!) If $\tau$ decays to both dark matter and the Standard Model, it can reheat the universe above the BBN scale to reconnect with the standard cosmology. Since in this scenario the dark and baryonic sectors were never in thermal equilibrium, their relative temperatures have in principle a great deal more
freedom to evade the acoustic oscillation bounds, and thus the bounds on DDDM quantity and interaction strength can be less severe.

Immediately after reheating, the $\tau$ decay products will generally be highly relativistic, and thus the temperatures of the sectors will be related to their relative energy densities by $\rho \sim g_* T_{\text{RH}}^4$. If we wish to have a dark sector with a very low relative temperature to evade acoustic bounds ($\xi = T_{\text{D}}/T_{\text{CMB}} \sim 0.1$, with $T_D$ the dark photon temperature when the CMB formed), we require
\begin{equation}
\frac{\rho^{\text{RH}}_D}{\rho^{\text{RH}}_B} = \frac{g^{\text{RH}}_{*,D}}{g^{\text{RH}}_{*,B}} \left(\frac{T^{\text{RH}}_{D}}{T^{\text{RH}}_{B}} \right)^4 = \left(\frac{g^{\text{RH}}_{*,D}}{g^{\text{RH}}_{*,B}}\right)^{-1/3} \left(\frac{T_{D}}{T_{\text{CMB}}} \right)^4 \sim \left(\frac{5.5}{86.25}\right)^{-1/3} \times 10^{-4} \sim 2.5 \times 10^{-4}.
\end{equation}
Here we have taken $g^{\text{RH}}_{*,B}$ to be the number of degrees of freedom above the $b$ quark mass scale, i.e.~we have assumed that $T^{\text{RH}}_{B}$ is somewhere in the few GeV range. We have also used conservation of entropy to relate the temperature at reheating to the temperature when the CMB formed, i.e.~$g^{\text{RH}}_{*,B} \left(T^{\text{RH}}_B\right)^3 = g^{\text{CMB}}_{*,B} T_{\rm CMB}^3$ and similarly in the dark sector. At the time that the CMB formed, we assume that both the dark and visible plasmas consist only of photons (two degrees of freedom in each sector). If the reheating temperature is below the charm and tau mass threshold, but around the QCD scale, the target ratio of energy densities drops slightly, closer to $2.2 \times 10^{-4}$.

Of course, if there are more relativistic degrees of freedom in the dark sector, or reheating occurs earlier, $g_{*,D}$ and $g_{*,B}$ may take slightly different values. Still, it is clear that a low DDDM temperature relative to the CMB requires a low DDDM energy density at reheating, and thus a low $\tau \rightarrow$ DDDM branching ratio. However, we can still achieve a relatively large present-day DDDM mass density by increasing the asymmetry per decay $\epsilon$ ---
that is, although few $\tau$s decay to DDDM particles, those that do are more likely to favor matter over antimatter, leading to few annihilations and a higher matter-to-energy ratio than in the visible sector.

Now, if we have such a group of interacting dark matter particles, is it possible that we would be able to see them in direct detection experiments? Suppose the DDDM sector is linked to the Standard Model via an effective 4-fermion coupling of the form $\frac{1}{\Lambda^2} \overline{\chi} \chi \overline{q} q$, where $X$ is the ``dark proton'' and $q$ is a quark. As we saw in the previous section, in order to have an observable direct detection cross-section near the current limits
while simultaneously requiring the sectors to be decoupled at the time of reheating (so that their temperatures are independent), dark matter must decouple at around $1$ GeV and $\tau$ must decay some time between then and the time of BBN at $10$ MeV. If the reheating temperature is close to the thermal decoupling temperature, we will also generate dark matter from the scattering of standard model particles (e.g.~$q{\overline q} \to \chi{\overline \chi}$). However, because the rate for such interactions goes like $T_{\rm SM}^5/\Lambda^4$, this is inefficient if the reheating temperature is below the decoupling temperature.

In short, as we will see, a model of interacting, observable dark matter produced through the non-thermal decay of long lifetime relic particles is indeed able to evade the pitfalls of its thermal cousin. The key ingredients of such a stratagem are a low energy density at the time of reheating, a high asymmetry factor for the decay, and a reheating temperature of $10 \text{ MeV}-1 \text{ GeV}$.

\subsection{Late time decay of weakly-coupled particles}

Before explaining the specific details of the model, let us remind the reader of some general features of late decays of particles like moduli. We parametrize the decay rate of the modulus $\tau$ as
\begin{equation}
\Gamma_\tau = \frac{1}{2\pi} \frac{m_\tau^3}{M_*^2}
\end{equation}
where $M_*$ is some ultraviolet scale near the Planck mass or perhaps slightly lower. The $\tau$s will decay roughly when their lifetime is around the Hubble time, that is, when $H \simeq \Gamma_\tau$. The energy previously contained in the large $\tau$ masses will be dumped into standard model particles, reheating the universe to a temperature of
\begin{equation}
\label{Trh}
T_{\text{RH}} \sim g_*^{-1/4} \Gamma_\tau^{1/2} M_p^{1/2} \simeq (5 \text{ MeV}) \left(\frac{10.75}{g_*}\right)^{1/4} \left(\frac{m_\tau}{100 \text{ TeV}}\right)^{3/2} \left(\frac{M_p}{M_*}\right)
\end{equation}
Equation \ref{Trh} comes from simply equating the energy densities, $\rho \sim H^2 M_p^2 \sim g_* T_{\rm RH}^4$.

From this equation, we see that for moduli masses $m_\tau \sim 100 \text{ --- } 1000 \text{ TeV}$ the moduli will reheat the universe before the epoch of BBN but at a temperature below the 1 GeV bound we established in the previous section to keep the baryonic and interacting DM sectors from thermally coupling.

Moduli decays can produce a large amount of entropy. If a modulus of mass 100 TeV leads to a reheating temperature of 5 MeV, then (after a brief time for the decay products to thermalize) any individual quantum of the modulus field must have given rise to a very large number of visible particles like photons, each of typical energy near $T_{\rm RH}$. Said differently: comparing the energy stored in the modulus before the decay and in radiation afterwards, we schematically have $\rho \sim m_\tau n_\tau \sim T_{\rm RH}^4$, so the entropy $s \sim T_{\rm RH}^3$ afterward is larger by a factor of about $m_\tau/T_{\rm RH}$ than the number of particles $n_\tau$ before the decay. Any pre-existing baryon number (or analogous asymmetry in the dark sector) is diluted after the decay by the factor 
\begin{equation}
Y_\tau = n_\tau/s_{\rm RH} = 3 T_{\rm RH}/4 m_\tau,
\end{equation}
which we will refer to as the ``dilution factor'' of the decay. For the case of a 100 TeV modulus and a reheating temperature of 5 MeV, we find $Y_\tau \approx 4 \times 10^{-8}$.

In our case, the modulus will decay to both the visible sector and the dark sector, so the energy density after the decay is schematically $\rho \sim T_{\rm vis}^4 + T_{\rm dark}^4$. However, we are interested precisely in the scenario with $T_{\rm vis}^4 \gg T_{\rm dark}^4$, so it is still appropriate to speak of a dilution factor where $T_{\rm RH}$ is interpreted as the reheating temperature {\em of the visible sector}. Asymmetries can either be produced during the decay process itself, or can have arisen from earlier physics and then been diluted by the factor $Y_\tau$.

\subsection{Baryogenesis from $\tau$ decay}

The small value of $Y_\tau$ is suggestive: it is only a couple of orders of magnitude larger than the tiny baryon-to-photon ratio $(n_B-n_{\overline B})/s \sim 10^{-10}$ that any model of baryogenesis seeks to explain. This fact was exploited by \cite{allahverdi2,allahverdi1} to build a model of baryogenesis via moduli decays. We will follow their lead as an example toy model, populating the baryonic and cold dark matter (CDM) sectors from the non-thermal decay of a string modulus $\tau$. Let us first review the baryogenesis mechanism for just the visible sector.

\begin{figure}
\begin{tikzpicture}[node distance=1cm and 1.3cm]
\coordinate[] (e1);
\coordinate[left=of e1,label=left:$N_\alpha$] (e2);
\coordinate[above right=of e1,label=right:$u^c$](e3);
\coordinate[below right=of e1,label=right:$X$](e4);

\draw[scalar] (e2) -- (e1);
\draw[electron] (e3) -- (e1);
\draw[electron] (e4) -- (e1);

\end{tikzpicture}
\begin{tikzpicture}[node distance=1cm and 1.3cm]

\coordinate[] (e1);
\coordinate[left=of e1, label=left:$N_\alpha$] (e2);
\coordinate[above right=of e1](e3);
\coordinate[below right=of e1](e4);
\coordinate[right=of e3,label=right:$u^c$](e5);
\coordinate[right=of e4,label=right:$X$](e6);

\draw[scalar] (e2) -- (e1);
\draw[electron] (e1) -- (e4) node [midway,below left] {$u^c$};
\draw[scalar] (e4) -- (e3) node [midway,right] {$N_\beta$};
\draw[electron] (e5) -- (e3);
\draw[electron] (e1) -- (e3) node [midway, above left] {$X$};
\draw[electron] (e6) -- (e4);

\end{tikzpicture}
\begin{tikzpicture}[node distance=1cm and 1.3cm]

\coordinate[] (e1);
\coordinate[left=of e1, label=left:$N_\alpha$] (e2);
\coordinate[right=of e1] (e3);
\coordinate[right=of e3] (e4);
\coordinate[above right=of e4,label=right:$u^c$] (e5);
\coordinate[below right=of e4,label=right:$X$] (e6);
\coordinate[above right=0.9 of e1,label=above:$X$] (n1);
\coordinate[below right=0.9 of e1,label=below:$u^c$] (n2);

\draw[scalar] (e2) -- (e1);
\semiloop[electron]{e3}{e1}{180};
\semiloopback[electron]{e3}{e1}{180};
\draw[scalar] (e3) -- (e4) node [midway, above] {$N_\beta$};
\draw[electron] (e5) -- (e4);
\draw[electron] (e6) -- (e4);
\end{tikzpicture}
\caption{Tree-level, vertex, and self-energy diagrams for the decay $N_\alpha \rightarrow X^* u^{c*}$ as in \cite{allahverdi2,allahverdi1}.}
\label{fig:baryoninterference}
\end{figure}
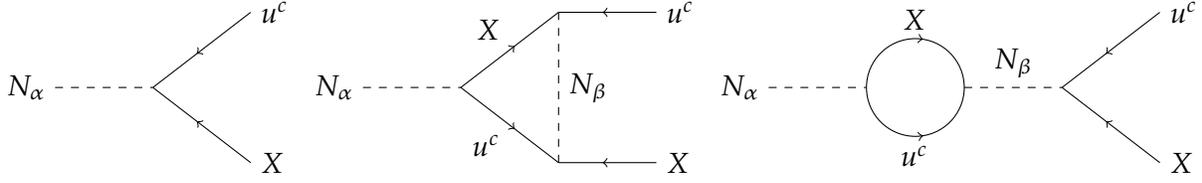

The model of \cite{allahverdi2} achieves baryogenesis by expanding the visible sector by introducing, in addition to the MSSM, two color singlets $N_\alpha$ (with $\alpha$ a flavor index) and a single flavor of colored triplets $X$, $\overline{X}$ with hypercharges $4/3, -4/3$ \cite{allahverdi1}. These additional fields have superpotential 
\begin{equation}
W_{\text{extra}} = \lambda_{i\alpha} N_\alpha u_i^c X + \lambda'_{ij} d_i^c d_j^c \overline{X} + \frac{M_\alpha}{2} N_\alpha N_\alpha + M_X X \overline{X}   \label{eq:BNVsuperpotential}
\end{equation}
We then assume that the $\tau$s have some decay rate into the scalar component of $N_\alpha$. $N_\alpha$ then itself decays through the diagrams in Figure \ref{fig:baryoninterference} to quarks and $X$s, which can in turn decay further to Standard Model particles. At tree-level, these diagrams and their hermitian conjugates have equal amplitudes, leading to equal production of matter and antimatter. However, as in many
common baryogenesis scenarios, interference terms between tree- and 1-loop-level diagrams will have different amplitudes, due to the relative factor of $i$ when the loop integral goes on shell \cite{Nanopoulos:1979gx,Fukugita:1986hr}. This leads to a disparity between the matter and anti-matter channels characterized by an asymmetry factor $\epsilon_B$ (see appendix \ref{sec:asymdetails} for the exact kinematic factors):
\begin{equation}
\epsilon_B = \frac{\Gamma(N_\alpha \rightarrow X u_i^c) - \Gamma(N_\alpha \rightarrow \overline{X} \overline{u_i^c})}{\Gamma(N_\alpha \rightarrow X u_i^c) + \Gamma(N_\alpha \rightarrow \overline{X} \overline{u_i^c})} \sim \sum_{i, j, \beta} \text{ Im}\left(\lambda_{i\alpha}\lambda^*_{i\beta} \lambda^*_{j\beta} \lambda_{j\alpha}\right)
\end{equation}
We assume that the scalar $N_\alpha$ decays to the fermions $u^c$ and $X$, whereas the scalar components of the supermultiplets obtain large SUSY-breaking masses and do not contribute to the asymmetry.

The overall baryon asymmetry is generally expressed in terms of $\eta_B$, the ratio of the net baryon number (baryons - anti-baryons) and the entropy density $s$. That is, we have
\begin{equation}
\eta_B = \frac{n_B - n_{\overline{B}}}{s} = Y_\tau \text{ Br}_N \epsilon_B
\end{equation}
where $Y_\tau$ is the dilution factor discussed above and $\text{Br}_N$ is the branching ratio to $N$. For the range of parameters we are considering, the dilution factor can be as low as $10^{-8}$ or lower; essentially, the huge release of energy when the $\tau$s decay is mostly carried as kinetic energy by the decay products, which is then dumped into symmetric matter and gauge bosons as the products thermalize, pumping up the entropy density and ``diluting'' the small number of asymmetrically produced particles. This factor by far dominates the observed asymmetry,
so that it is easily possible to get the observed value of $\eta_B \sim 10^{-10}$ with reasonable values of $\text{Br}_N$ and $\epsilon_B$, for example with $T_{RH} \sim 100 \text{ MeV}$, $m_\tau \sim 100 \text{ TeV}$, $\text{Br}_N \sim 0.01$, $\epsilon_B \sim 0.01$.

In such a model, it is reasonable to assume that cold dark matter, whatever form it may take, is also produced from the $\tau$ decay, since the heavy dilution factor will render any pre-existing dark matter irrelevant. If each decay produces a single CDM particle $\chi$, their number density is then given by
\begin{equation}
\label{CDM}
\frac{n_\chi}{s} = Y_\tau \text{ Br}_\chi
\end{equation}

As discussed in \cite{allahverdi1}, for reasonable values of the dark matter annihilation cross-section ($\langle \sigma_{\text{ann}} v \rangle < 10^{-33} \text{ cm}^2$) the non-thermally produced CDM does not experience significant annihilation after modulus decay, and so equation \ref{CDM} gives the final CDM abundance.

Thus, the baryon and CDM density ratios are given by
\begin{equation}
\frac{\Omega_\text{B}}{\Omega_{\text{DM}}} \simeq \frac{1 \text{ GeV}}{m_\chi} \frac{\epsilon_B \text{ Br}_N}{\text{ Br}_\chi}
\end{equation}
and so if the branching ratios are of an order the correct CDM abundance can be obtained for CDM masses around 0.05 GeV with $\epsilon_B \sim 0.01$.

\subsection{Baryon number violation at low energies}

This model of baryogenesis involves new sources of baryon number violation near the TeV scale, so it is important to check that it is consistent with constraints. Because the new interactions we postulate violate baryon number but not lepton number (and we have not added any new light fermions that the proton can decay to), there is no new source of proton decay in this model. However, neutron--antineutron oscillations and dinucleon decay processes arise, so we must check that their rate is consistent with observation. The case of neutron--antineutron oscillations was briefly mentioned in \cite{allahverdi2} with the conclusion that either flavor-violating couplings or mild degeneracies in the spectrum could alleviate the bound. However, dinucleon decay is also an important constraint that must be checked.

Notice that all four terms in the superpotential (\ref{eq:BNVsuperpotential}) are necessary to violate baryon number. For instance, if we set $M_\alpha = 0$, we could assign $B_X = -B_{\overline X} = -2/3$ and $B_N = +1$. But then adding a nonzero mass term $M_\alpha N_\alpha N_\alpha$ violates baryon number by two units. Soft supersymmetry violating holomorphic scalar mass terms $b_{N\alpha} N_\alpha N_\alpha$ and $b_X X {\overline X}$ can also contribute to baryon number violating processes.

\begin{figure}
\centering
\begin{tikzpicture}[line width=1.5 pt]
    \draw[majfermion] (0,1)--(1,1);
    \draw[majfermion] (1,1)--(1,0);
    \draw[majfermion] (1,0)--(0,0);
    \draw[scalar] (1,0)--(1,-1);
    \draw[majfermion] (0,-1)--(1,-1);
    \draw[majfermion] (1,-1)--(2,-1);
    \draw[majfermion] (2,-1)--(3,-1);
    \draw[scalar] (2,-1)--(2,0);
    \draw[majfermion] (2,0)--(3,0);
    \draw[majfermion] (2,0)--(2,1);
    \draw[majfermion] (2,1)--(3,1);
    \draw[scalar] (2,1)--(1,1);
    \node[cross] at (1,-0.5) {};
    \node[cross] at (2, -0.5) {};
    \node at (-0.2, 1) {$u$};
    \node at (-0.2, 0) {$d$};
    \node at (-0.2, -1) {$d$};
    \node at (3.2, 1) {$\overline{u}$};
    \node at (3.2, 0) {$\overline{d}$};
    \node at (3.2, -1) {$\overline{d}$};
    \node at (1.5, 1.2) {$N$};
    \node at (1.5, -1.3) {$\tilde{g}$};
    \node at (0.7, 0.5) {$X$};
    \node at (2.3, 0.5) {$X$};
    \node at (0.7, -0.25) {$\tilde{b}$};
    \node at (0.7, -0.75) {$\tilde{d}$};
    \node at (2.3, -0.25) {$\tilde{b}$};
    \node at (2.3, -0.75) {$\tilde{d}$};
\end{tikzpicture}
\caption{$\Delta B = 2$ diagram which gives rise to possible neutron-antineutron oscillation}
\label{nbridge}
\end{figure}
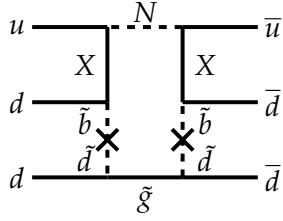

We first consider $n$--$\overline{n}$ oscillations. The color contraction of the $ddX$ term requires that $\lambda'_{ij}$ be antisymmetric, so any diagram necessarily involves down-type quarks of different flavors. The leading contribution appears at one loop, as in Figure \ref{nbridge}. This diagram requires the presence of flavor changing neutral currents through off-diagonal terms in the right-handed squark mass matrix, $\Delta^{{\tilde d}RR}_{ij}$.  The size of these contributions is necessarily suppressed if squarks and gluinos are near the TeV scale, due to FCNC constraints. For instance, if squarks and gluinos have mass 1 TeV we have $\left|{\rm Re}(\Delta^{{\tilde d}RR}_{ds})\right| \lesssim  6 \times 10^{-2} m_{{\tilde d},{\tilde s}}^2$ \cite{Ciuchini:1998ix}. The bound becomes somewhat stronger if there is flavor violation in both the left- and right-handed sectors. However, if we take the squarks to be heavier, as in mildly split supersymmetry, the bounds weaken. Bounds when the third generation squarks are involved are also weaker. So we expect a modest, but not necessarily extreme, suppression of $n$--$\overline{n}$ oscillations in theories consistent with FCNC bounds. Specific flavor models (like \cite{csaki}) might predict a more significant suppression.

Integrating out the heavy particles in Fig.~\ref{nbridge} generates a six-fermion (dimension-9) operator of the schematic form
\begin{equation}
{\cal L}_{\Delta B} \sim \frac{\lambda^2 \lambda'^2 g_s^2}{16 \pi^2} \left(\frac{\Delta_{db}}{m_{\tilde d}^2 m_{\tilde b}^2}\right)^2 \frac{b_N}{m_{\tilde N}^4} m_{\rm scalar} (u^c d^c d^c)^2 + {\rm h.c.}.
\end{equation}
Here $m_{\rm scalar}$ is a mass scale resulting from doing the full loop integral; if we assume that the scalars are at least modestly heavier than the fermions, i.e.~$m_{\tilde q}, m_{\tilde N} \gtrsim m_X, m_{\tilde g}$, then $m_{\rm scalar}$ is roughly expected to be set by the smallest of the scalar mass scales. This operator built out of quarks must be matched onto nucleon states at the QCD scale to produce an effective operator ${\cal L}^{\rm eff}_{\Delta B} = m_{\rm osc} n n + {\rm h.c.}$. The matrix elements are rather uncertain; we will follow the recent review \cite{Phillips:2014fgb} and assume
\begin{equation}
\left<n \middle| (u^c d^c d^c)^2 \middle| n\right> \sim 3 \times 10^{-5}~{\rm GeV}^6,
\end{equation}
although there is an order-one uncertainty in existing lattice calculations \cite{Buchoff:2012bm}. Using this we have the order-of-magnitude estimate:
\begin{equation}
t_{\rm osc} = m_{\rm osc}^{-1} \sim 2.9 \times 10^{8}~{\rm s}~\left(\frac{10^{-1}}{\lambda}\right)^2 \left(\frac{10^{-1}}{\lambda'}\right)^2 \left(\frac{M_{\rm eff}}{25~{\rm TeV}}\right)^5,    \label{eq:neutronantineutron}
\end{equation}
with $M_{\rm eff}^5 = m_{\tilde d}^4 m_{\tilde b}^4 m_{\tilde N}^4/(\Delta_{db}^2 b_N m_{\rm scalar})$ a combination of the different scalar mass scales appearing in the diagram. This is to be compared with the experimental bound \cite{kamiokande}
\begin{equation}
t_{\rm osc} \geq 2.7 \times 10^8~{\rm s}.
\end{equation}
From this estimate we see that we are safe from constraints provided that the average mass scale is above about 25 TeV for couplings of around 0.1. In fact, it is reasonable to take $\Delta_{db} \ll m_{\tilde d, \tilde b}^2$ and also to assume some flavor structure in the $\lambda$ and $\lambda'$ couplings that suppresses them for the first generation quarks, so our model is very safe from nucleon oscillation bounds.

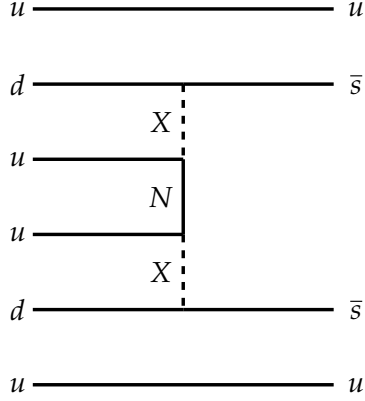
\begin{figure}
\centering
\begin{tikzpicture}[line width=1.3 pt]
    \draw[majfermion] (0, 2.5)--(4,2.5);
    \draw[majfermion] (0, 1.5)--(2,1.5);
    \draw[majfermion] (2, 1.5)--(4,1.5);
    \draw[scalar] (2, 1.5)--(2,0.5);
    \draw[majfermion] (0, 0.5)--(2,0.5);
    \draw[majfermion] (2, 0.5)--(2,-0.5);
    \draw[majfermion] (0, -0.5)--(2,-0.5);
    \draw[scalar] (2, -0.5)--(2, -1.5);
    \draw[majfermion] (0, -1.5)--(2,-1.5);
    \draw[majfermion] (2, -1.5)--(4,-1.5);
    \draw[majfermion] (0, -2.5)--(4,-2.5);
    \node at (-0.2,2.5) {$u$};
    \node at (-0.2,1.5) {$d$};
    \node at (-0.2,0.5) {$u$};
    \node at (-0.2,-2.5) {$u$};
    \node at (-0.2,-1.5) {$d$};
    \node at (-0.2,-0.5) {$u$};
    \node at (4.3, 2.5) {$u$};
    \node at (4.3, 1.5) {$\overline{s}$};
    \node at (4.3, -1.5) {$\overline{s}$};
    \node at (4.3, -2.5) {$u$};
    \node at (1.7, 1) {$X$};
    \node at (1.7, -1) {$X$};
    \node at (1.7, 0) {$N$};
\end{tikzpicture}
\caption{Leading diagram contributing to the dinucleon decay $pp\rightarrow K^+ K^+$}
\label{dinuc}
\end{figure}

Another potentially important constraint comes from dinucleon decay, $p p \rightarrow K^+ K^+$. Unlike the case of neutron--antineutron oscillation, here the process intrinsically changes flavor. As a result, this process occurs at tree level as illustrated in Figure \ref{dinuc}. The effective Lagrangian is schematically 
\begin{equation}
{\cal L}_{\Delta B} \sim \frac{\lambda^2 \lambda'^2 b_{X\overline{X}}^2}{m_N m_{\tilde X}^4 m_{\tilde {\overline X}}^4} \left(u^c d^c s^c\right)^2 + {\rm h.c.},
\end{equation}
which again must be matched to a hadronic matrix element $\left<K^+K^+\middle| \left(u^c d^c s^c\right)^2 \middle| pp\right>$. Following \cite{goity} and \cite{csaki}, we approximate the hadronic physics with powers of ${\tilde \Lambda} \approx 150~{\rm MeV}$ and estimate
\begin{align}
\tau &\sim \frac{m_p^2 }{128 \pi \rho_{\rm nuc} {\tilde \Lambda}^{10}} \left(\frac{m_N m_{\tilde X}^4 m_{\tilde {\overline X}}^4}{\lambda_u^2 \lambda'^2_{ds} b_{X\overline{X}}^2}\right)^2 \nonumber \\
&\sim 4.5 \times 10^{33}~{\rm yrs}~\left(\frac{10^{-2}}{\lambda_u}\right)^2 \left(\frac{10^{-2}}{\lambda'_{ds}}\right)^2\left(\frac{m_N}{5~{\rm TeV}}\right)^2 \left(\frac{m_{\tilde X}}{40~{\rm TeV}}\right)^{8} \left(\frac{m_{\tilde {\overline X}}}{40~{\rm TeV}}\right)^{8} \left(\frac{(10~{\rm TeV})^2}{b_{X \overline{X}}}\right)^4.    \label{eq:dinucleon}
\end{align}
with $m_p$ the proton mass and $\rho_{\rm nuc}  \sim 0.25~{\rm fm}^{-3}$ the approximate number density of nucleons in nuclear matter. This is to be compared to the experimental bound $\tau \geq 1.7 \times 10^{32} \text{ yrs}$ \cite{litos}.

Given our results we can see that the dinucleon decay constraint (\ref{eq:dinucleon}) is more stringent than the neutron--antineutron oscillation constraint (\ref{eq:neutronantineutron}). It requires us to take a fermion, $N$, to be heavy, rather than just the scalars. It also cannot be suppressed with assumptions about the flavor structure of the squark mass matrix. It can, however, be suppressed by flavor structure in the couplings $\lambda_i$ and $\lambda'_{ij}$, as well as if the $b$-term contribution to the $X, {\overline X}$ mass matrix is somewhat small compared to the diagonal soft mass terms. Furthermore, because $\tau(pp \to K^+ K^+)$ depends on such large powers of the scalar mass scales, making them only modestly heavier can help to evade the bound. (Conversely, if the nuclear scale ${\tilde \Lambda}$ is larger than expected, this can dramatically strengthen the bound and force us to consider heavier scalars.)

Discussing a detailed model of flavor physics is well beyond the scope of this paper. Notice that $\lambda_i$ transforms in the fundamental of SU(3)$_u$ and $\lambda'_{ij}$ transforms as an antisymmetric tensor of SU(3)$_d$; as a result, these are entirely different flavor-breaking spurions from the Yukawa matrices $Y_u$ and $Y_d$, and we cannot assume Minimal Flavor Violation. In many cases flavor constraints can be satisfied with smaller symmetries \cite{Barbieri:2012uh}, and large scalar masses can help. We imagine that the most likely ansatz is that the $\lambda$ and $\lambda'$ couplings involving the third generation are largest, and the others are suppressed. (This structure might arise, for instance, from extending a model like \cite{Heidenreich:2014jpa}.) In that case, the bounds (\ref{eq:dinucleon}) and (\ref{eq:neutronantineutron}) will be easily satisfied because first-generation quarks appear, whereas the larger third-generation couplings will play the key role in the generation of a baryon asymmetry from the modulus decay.

\subsection{DDDM from $\tau$ decay}

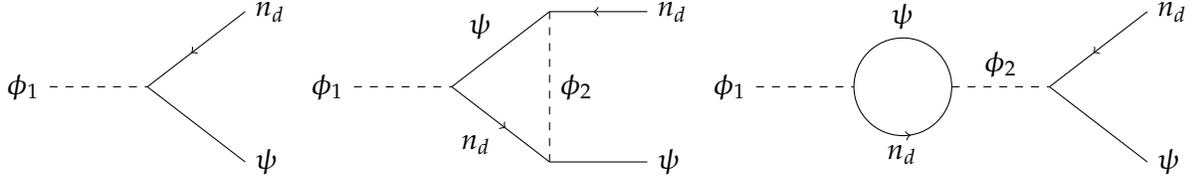
\begin{figure}
\begin{tikzpicture}[node distance=1cm and 1.3cm]
\coordinate[] (e1);
\coordinate[left=of e1,label=left:$\phi_1$] (e2);
\coordinate[above right=of e1,label=right:$n_d$](e3);
\coordinate[below right=of e1,label=right:$\psi$](e4);

\draw[scalar] (e2) -- (e1);
\draw[electron] (e3) -- (e1);
\draw[majorana] (e4) -- (e1);

\end{tikzpicture}
\begin{tikzpicture}[node distance=1cm and 1.3cm]

\coordinate[] (e1);
\coordinate[left=of e1, label=left:$\phi_1$] (e2);
\coordinate[above right=of e1](e3);
\coordinate[below right=of e1](e4);
\coordinate[right=of e3,label=right:$n_d$](e5);
\coordinate[right=of e4,label=right:$\psi$](e6);

\draw[scalar] (e2) -- (e1);
\draw[electron] (e1) -- (e4) node [midway,below left] {$n_d$};
\draw[scalar] (e4) -- (e3) node [midway,right] {$\phi_2$};
\draw[electron] (e5) -- (e3);
\draw[majorana] (e1) -- (e3) node [midway, above left] {$\psi$};
\draw[majorana] (e4) -- (e6);

\end{tikzpicture}
\begin{tikzpicture}[node distance=1cm and 1.3cm]

\coordinate[] (e1);
\coordinate[left=of e1, label=left:$\phi_1$] (e2);
\coordinate[right=of e1] (e3);
\coordinate[right=of e3] (e4);
\coordinate[above right=of e4,label=right:$n_d$] (e5);
\coordinate[below right=of e4,label=right:$\psi$] (e6);
\coordinate[above right=0.9 of e1,label=above:$\psi$] (n1);
\coordinate[below right=0.9 of e1,label=below:$n_d$] (n2);

\draw[scalar] (e2) -- (e1);
\semiloop[electron]{e3}{e1}{180};
\semiloop[majorana]{e1}{e3}{0};
\draw[scalar] (e3) -- (e4) node [midway, above] {$\phi_2$};
\draw[electron] (e5) -- (e4);
\draw[majorana] (e4) -- (e6);
\end{tikzpicture}
\caption{Diagrams for the proposed $\phi_1 \rightarrow n_d\psi$ decay}
\label{fig:dddminterference}
\end{figure}

The DDDM sector can be produced in a similar manner to the Standard Model particles. Let us consider perhaps the simplest DDDM sector which can generate an asymmetric population from modulus decay. We introduce chiral superfields $\psi$, $\phi_1$, $\phi_2$, $n_d$, and ${\overline n}_d$ with superpotential
\begin{equation}
W = \sum_i \left(\lambda_i \phi_i \psi n_d + \frac{1}{2} M_i \phi_i^2\right) + \frac{1}{2} M_\psi \psi^2 + M_{n_d} n_d {\overline n}_d,
\end{equation}
and we take $M_1 > M_2$ without loss of generality. We focus on a scenario in which the modulus $\tau$ decays to the scalar components of $\phi_1$ and $\phi_2$ which in turn have out-of-equilibrium decays to the fermions in the $\psi$ and $n_d$ multiplet. The decay of the $\phi$s exhibits CP violation from the interference of the tree and 1-loop diagrams in Figure \ref{fig:dddminterference}. We assume that the fermion masses are all relatively small compared to the supersymmetry breaking scale, whereas the scalars in each multiplet acquire soft masses of order $m_{3/2}^2$, roughly of the same order of magnitude as the $\tau$ mass. For simplicity, we can assume that the scalar partners of $\psi$ and $n_d$ are either kinematically out of reach or have suppressed branching ratios, while the modulus frequently decays to the heavy scalars $\phi$.

In analogy with the previous computation, this interference generates an asymmetry factor $\epsilon_d \sim \sum_{j} \text{ Im}\left[\lambda_1\lambda_j^* \lambda_j^* \lambda_1\right]$. For more detail, see appendix \ref{sec:asymdetails}.

Suppose that $n_d$ then decays through some separate process to a dark ``proton,'' ``electron,'' and ``neutrino,'' in analogy to the weak interactions. That is, there is the additional term in the Lagrangian
\begin{equation}
\mathcal{L} = \frac{g}{\Lambda^2} \overline{p}_d n_d \overline{l}_d \nu_d + {\rm h.c.}
\end{equation}
(In ref.~\cite{Fan:2013yva} the dark proton was denoted $X$ and the dark electron $C$. Here we prefer a notation that makes the analogy to the Standard Model more explicit.) If the dark neutron is sufficiently heavy compared to the other dark particles, then the neutrons produced by modulus decay will further decay into equal populations of dark protons and electrons which can form into a modern-day population of dark atoms. Thus the asymmetry of the $\phi$ decays is converted directly into a ``dark baryon'' or DDDM asymmetry:
\begin{equation}
\eta_d = \frac{n_d - n_{\overline{d}}}{s} = Y_\tau Br_\phi \epsilon_d.
\end{equation}

\subsection{Requirements}
\label{subsec:requirements}

For the mechanism discussed above, we need the following features:
\begin{itemize}
\item A modulus that decays dominantly into Standard Model fields.
\item A subdominant modulus decay width to DDDM sector fields, ${\rm Br}_d \lesssim 2 \times 10^{-4}$.
\item A dilution factor times an asymmetry generation factor that can account for the baryonic abundance,
  \begin{equation}
      \epsilon_B \frac{T_{\rm RH}}{m_\tau} {\rm Br}_N \sim 10^{-10}.
  \end{equation}
\item A similar factor to account for the DDDM abundance, for which a reasonable target for detectability might be $\Omega_d \sim 0.1 \Omega_B$, i.e.~$(n_d-n_{\overline d})\sim 0.1 m_p (n_B - n_{\overline{B}})/m_{p_d}$:
  \begin{equation}
      \epsilon_d \frac{T_{\rm RH}}{m_\tau} {\rm Br}_\phi \sim 10^{-11} \frac{m_p}{m_{p_d}}.
  \end{equation}   
  In this case ${\rm Br}_\phi \lesssim {\rm Br}_d \lesssim  2 \times 10^{-4}$.
\end{itemize} 
A set of values that achieves these requirements is:
\begin{align}
\epsilon_B \sim 10^{-3} \quad \quad \epsilon_d \sim 10^{-2} \quad \quad M_* \sim 4 \times 10^{16}~{\rm GeV} \nonumber \\
m_\tau \sim 100~{\rm TeV} \quad \quad T_{\rm RH} \sim 300~{\rm MeV} \quad \quad Y_\tau \sim 3 \times 10^{-6} \nonumber \\
2 {\rm Br}_\phi \sim {\rm Br}_d \sim 2 \times 10^{-4} \quad \quad {\rm Br}_N \sim 3 \times 10^{-2} \quad \quad m_{p_d} \sim 3 m_p.
\end{align}
First, note that this would not work with a modulus that has only Planck-suppressed couplings: in that case, for a given $m_\tau$, the reheating temperature would be smaller, and we would require uncomfortably large values of either asymmetry parameters $\epsilon$ or branching ratios. 

However, the value of $M_*$ that works is quite close to the GUT scale, so it plausibly connects to well-motivated high-scale physics. The modulus mass $m_\tau \sim 100~{\rm TeV}$ fits comfortably with the scalar masses in the tens of TeV range that we found worked well for avoiding bounds from low-energy baryon number violation constraints, and also with theories of mini-split supersymmetry in which gauginos may be near the TeV scale. We have $\epsilon_B \sim 10^{-3}$ and $\epsilon_d \sim 10^{-2}$, which is a sensible result for ${\cal O}(1)$ values of the superpotential couplings given the factors of $24\pi$ and $8\pi$ appearing in the denominators (see appendix \ref{sec:asymdetails}).

The one number here that may seem mildly uncomfortable is the total branching ratio to the dark sector, ${\rm Br}_d \sim 2 \times 10^{-4}$, which was chosen to accommodate the constraint that the dark photon temperature at the time the CMB formed should be no more than about a tenth of the photon temperature in the regime with interesting interactions, as discussed in \S\ref{sec:tension}. If the modulus decayed democratically to all species, we might expect this to be closer to a few percent, so we require the modulus couplings to dark sector degrees of freedom to systematically be about an order of magnitude smaller than its couplings to visible sector degrees of freedom. The existence of modulus--baryon couplings and baryon--dark sector couplings inevitably induces modulus--dark sector couplings through loops, but these will contribute branching fractions safely below the $10^{-4}$ level we require, so this assumption is radiatively stable. The whole story hangs together rather nicely.

Let us pause here to mention some other possibilities for achieving the dark matter and dark radiation abundance.

{\bf \em Diluting a pre-existing asymmetry.} Although we assume that the moduli overwhelmingly dominate the {\em energy} density of the universe before they decay, they might not be so dominant in terms of the {\em number} density. In particular, suppose that before the moduli decay, we have baryon and dark asymmetries at the level $(n_B - n_{\overline{B}})/n_\tau \sim 10^{-2}$ and $(n_d-n_{\overline d})/n_\tau \sim 10^{-3}$. In that case, even if the modulus decays are completely symmetric, a dilution factor $Y_\tau \sim 10^{-8}$  can lead to the small asymmetries of interest in the late-time universe. We have now shifted the model-building problem to earlier times: why are there relatively large asymmetries in place before the moduli decay? A large toolkit of baryogenesis models is available for formulating possible answers.

{\bf \em Freeze-in of dark radiation.} In the above discussion, we have focused on dark radiation generated directly from decays of moduli. However, even if the modulus only coupled to the Standard Model fields, we know that if its decays reheated the visible sector above a GeV we would generate a large amount of dark radiation from thermalization of the visible and dark sectors. At lower reheating temperatures, the dark and visible sectors never thermalize, but we still produce dark particles from scattering of visible particles, so the dark radiation abundance will gradually increase before ``freezing in'' \cite{Chung:1998rq,Hall:2009bx}. Depending on the precise reheating temperature, this in itself could account for an interesting but allowed level of dark radiation.

\section{Further remarks on modeling}
\label{section:smallnumbers}

In this section we will make some brief remarks on other aspects of model-building. First, we comment on the fact that the modulus must have smaller couplings to the dark sector than to the visible sector, despite the fact that we want direct dark--visible couplings for direct detection, so that one might think there is no fundamental distinction between sectors. This could be a modest accident or could arise from a more complex model. Second, we comment on constraints from kinetic mixing, which can be safely satisfied.

\subsection{Remarks on the size of couplings}

The couplings listed in \S\ref{subsec:requirements} involve no exceptionally large hierarchies to explain. We have not yet explained how direct detection happens; one possibility is a $Z'$ boson with a mass near the TeV scale mediating interactions between the sectors.

One could attempt to go further and explain the modest suppression of modulus branching ratio to the dark sector, $\lesssim 2 \times 10^{-4}$, via modulus couplings that are geometrically suppressed. If a very mildly warped extra dimension has an IR scale $M_* \approx M_{\rm GUT}$, the modulus is a radion, visible fields live on the IR brane, and dark sector fields live on the UV brane, this could help explain both why the modulus couplings to the visible sector are stronger than gravitational \cite{Randall:1999ee,Charmousis:1999rg,Csaki:1999mp} and why the decays to dark sector fields are suppressed. The radion could be light due to SUSY-breaking stabilization mechanisms \cite{Luty:2000ec}. The $Z'$-mediated direct detection process would be unsuppressed because massless gauge fields have flat profiles in the extra dimension, and the $Z'$ mass could be set by SUSY breaking to be well below the compactification scale.

These elaborations are not necessary for our model to work, but if there are multiple hidden sectors localized in different places, similar ideas may help to explain why our universe has relatively low amounts of dark radiation due to much more suppressed reheating of some of the sectors.
 
\subsection{Kinetic mixing}

One concern about the specific $Z'$ model we have outlined is that the $Z'$ interacting with both the visible and dark sectors could lead to a kinetic mixing that is excluded by experimental limits \cite{Davidson:2000hf}. As discussed in \cite{Fan:2013yva}, if we want to consider dissipative dynamics in the dark sector that lead to dark matter cooling phenomena in galaxies, we need to rely on a ``dark electron'' mass below about an MeV, for which the kinetic mixing parameter must be $\lesssim 10^{-9}$. Kinetic mixing of U(1)s is a marginal operator and so can be generated at any scale. To avoid generating an unacceptably large mixing from heavy particles (e.g.~string states), we must imagine that at least one of hypercharge or the dark sector U(1) is embedded in a nonabelian group, so that in the ultraviolet the kinetic mixing is a higher-dimension operator.

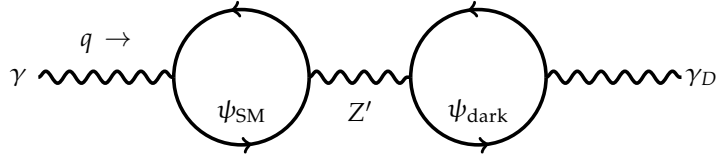
\begin{figure}[!h]
\begin{center}
\scalebox{0.9}{
\begin{tikzpicture}[line width=1.5 pt]
\draw[photon] (-2.0,0.0)--(0.0,0.0);
\node at (-1.0,0.5) {$q\, \rightarrow$};
\draw[electron] (2.0,0.0) arc(0:180:1.0);
\draw[electron] (0.0,0.0) arc(180:360:1.0);
\node at (1.0,-0.5) {$\psi_{\rm SM}$};
\draw[photon] (2.0,0.0)--(3.5,0.0);
\node at (2.75,-0.5) {$Z'$};
\draw[electron] (5.5,0.0) arc(0:180:1.0);
\draw[electron] (3.5,0.0) arc(180:360:1.0);
\node at (4.5,-0.5) {$\psi_{\rm dark}$};
\draw[photon] (5.5,0.0)--(7.5,0.0);
\node at (7.8,0.0) {$\gamma_D$};
\node at (-2.3,0.0) {$\gamma$};
\end{tikzpicture}
}
\end{center}
\caption{Two-loop kinetic mixing with intermediate $Z'$. Because this diagram is not 1PI, it does not generate a renormalization of the mixing above the $Z'$ mass scale. When the $Z'$ is integrated out, a suppressed mixing of order $q^2/m_{Z'}^2$ (a higher-dimension operator) is generated.} 
\label{fig:kineticmixing2loop}
\end{figure}

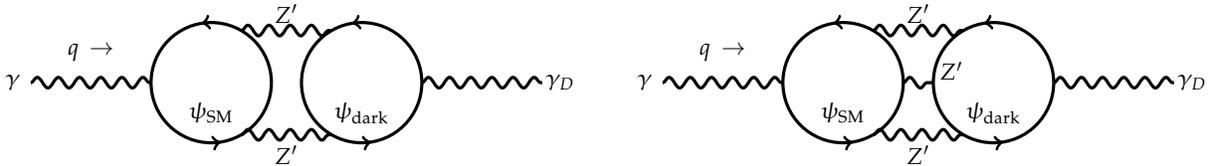
\begin{figure}[!h]
\begin{center}
\scalebox{0.8}{
\begin{tikzpicture}[line width=1.5 pt]
\draw[photon] (-2.0,0.0)--(0.0,0.0);
\node at (-1.0,0.5) {$q\, \rightarrow$};
\draw[electron] (2.0,0.0) arc(0:180:1.0);
\draw[electron] (0.0,0.0) arc(180:360:1.0);
\draw[photon] (1.5,0.867)--(3.0,0.867);
\node at (2.25,1.15) {$Z'$};
\node at (2.25,-1.2) {$Z'$};
\draw[photon] (1.5,-0.867)--(3.0,-0.867);
\node at (1.0,-0.5) {$\psi_{\rm SM}$};
\draw[electron] (4.5,0.0) arc(0:180:1.0);
\draw[electron] (2.5,0.0) arc(180:360:1.0);
\node at (3.5,-0.5) {$\psi_{\rm dark}$};
\draw[photon] (4.5,0.0)--(6.5,0.0);
\node at (6.8,0.0) {$\gamma_D$};
\node at (-2.3,0.0) {$\gamma$};
\begin{scope}[shift={(10.5,0.0)}]
\draw[photon] (-2.0,0.0)--(0.0,0.0);
\node at (-1.0,0.5) {$q\, \rightarrow$};
\draw[electron] (2.0,0.0) arc(0:180:1.0);
\draw[electron] (0.0,0.0) arc(180:360:1.0);
\draw[photon] (1.5,0.867)--(3.0,0.867);
\node at (2.25,1.15) {$Z'$};
\node at (2.25,-1.2) {$Z'$};
\draw[photon] (1.5,-0.867)--(3.0,-0.867);
\node at (1.0,-0.5) {$\psi_{\rm SM}$};
\draw[electron] (4.5,0.0) arc(0:180:1.0);
\draw[electron] (2.5,0.0) arc(180:360:1.0);
\node at (3.5,-0.5) {$\psi_{\rm dark}$};
\draw[photon] (4.5,0.0)--(6.5,0.0);
\node at (6.8,0.0) {$\gamma_D$};
\node at (-2.3,0.0) {$\gamma$};
\draw[photon] (2.0,0.0)--(2.5,0.0);
\node at (2.8,0.2) {$Z'$};
\end{scope}
\end{tikzpicture}
}
\end{center}
\caption{Three- and four-loop kinetic mixing through the intermediate $Z'$. The three-loop diagram vanishes because loops of particles and antiparticles cancel. The four-loop diagram is nonzero but can be safely small.} 
\label{fig:kineticmixing34loop}
\end{figure}

The massive $Z'$ we have added to the theory to couple the two sectors can lead to new, infrared contributions to kinetic mixing in the low-energy effective theory. (In particular, these are distinct from any effect discussed in the appendix of \cite{Fan:2013yva}, which did not postulate a massive bosonic mediator between the sectors.) The first concern might be the two-loop diagram in figure \ref{fig:kineticmixing2loop}, but this diagram is not 1PI and so does not enter in the renormalization group equations. It does lead to a higher-dimension operator when we integrate out the $Z'$, which parametrically scales as two loop factors times $q^2/m_{Z'}^2$, safely negligible at the $q^2$ values of cosmological interest. We can then consider the three- and four-loop diagrams of figure \ref{fig:kineticmixing34loop}, which potentially generate a genuine kinetic mixing from renormalization group evolution. However, the three-loop diagram cancels because we must sum both particles and antiparticles in the loop, as explained in the appendix of \cite{Fan:2013yva}. The four-loop diagram leads to a nonzero kinetic mixing, but it is of order $e e_D g_{Z'}^6/(16\pi^2)^4$, which can be sufficiently small to satisfy the cosmological constraint.

\section{Conclusions}
\label{sec:conclusions}

We have pointed out that if dark matter has significant interactions with dark radiation, there is a generic tension between having detectably large couplings of the Standard Model to dark matter without overproducing dark radiation in the early universe. Nonthermal physics at relatively late times, such as exit from a matter-dominated phase through particle decays that reheat the universe to a temperature between that of BBN and about 1 GeV, can evade this problem. We have discussed a model that can do this, as a proof of principle. There are still some further model-building issues we might concern ourselves with. For example, we have to embed the particles we have discussed in a GUT, to avoid dangerously large kinetic mixing from ultraviolet physics. We have not fully explored how supersymmetry fits into the picture or how the radion in our extra-dimensional picture (or more generally, the modulus in any UV completion of our scenario) is stabilized. We also have not explained the lightness of the dark electron mass, which is suggestive of a dark sector built from a chiral theory more closely resembling the Standard Model. However, we expect that resolution of these issues is largely orthogonal to the cosmological questions discussed in this paper.

The general sort of scenario that we have discussed here may also be relevant for other scenarios with dark matter coupled to dark radiation. Recently theories with nonabelian (self-interacting) dark radiation have gained some attention \cite{Jeong:2013eza,Blinov:2014nla,Buen-Abad:2015ova,Lesgourgues:2015wza,Choquette:2015mca}. We could consider cosmology driven by decaying moduli in this context, although new issues arise: as emphasized in \cite{Buen-Abad:2015ova}, dark matter--dark radiation interactions do not decouple in nonabelian theories as they do in abelian ones, and so cosmological bounds on the interaction strength can be much more stringent. Simply modifying thermal physics above the scale of BBN, as we have done in this paper, is insufficient to rescue such nonabelian theories from their constraints.

Recently a great deal of renewed interest has arisen in the Twin Higgs scenario, in which the mass of the Standard Model Higgs boson can receive extra protection by doubling the Standard Model fields \cite{Chacko:2005pe}. In such models, twin sector copies of light Standard Model fields can play the role of dark radiation. Because the Higgs boson must have equal couplings to the two sectors in order for the protection mechanism to work, this is a class of dark radiation models in which the two sectors unavoidably have a low thermal decoupling temperature. The late-time asymmetric reheating scenario that we have discussed could fit into supersymmetric extensions of the Twin Higgs scenario \cite{Chang:2006ra, Craig:2013fga} with moduli fields. The challenge in the Twin Higgs case will be to provide a better explanation of the asymmetric modulus branching ratios into the two sectors. The underlying theory must have an exact ${\mathbb Z}_2$ symmetry in order for the twin protection mechanism to operate. On the other hand, this symmetry must be spontaneously broken. The challenge is how to extend this broken symmetry structure to the moduli sector. Perhaps there are two separate moduli fields, one coupling more to the Standard Model and one coupling more to the twin sector, and one of the two fields dominates the energy density of the universe purely from an accident of initial conditions. We will not explore the application to the Twin Higgs in further detail in this paper, but it would be an interesting exercise to try to construct a fully realized model.

\section*{Acknowledgments}

We thank an anonymous referee for important comments. The authors are supported in part by the NSF Grant PHY-1415548.  MR is supported in part by the NASA Astrophysics Theory Grant NNX16AI12G and by the National Science Foundation under Grant No. PHYS-1066293 and the hospitality of the Aspen Center for Physics.

\appendix

\section{Asymmetry factors}
\label{sec:asymdetails}

This appendix will give explicit forms for the asymmetry factors $\epsilon$ in our toy models for baryogenesis and the creation of asymmetric DDDM. The asymmetry comes from the interference of tree and one-loop diagrams, and arises from a simple argument. Suppose we have some decay processes $\psi \rightarrow f$ and $\overline{\psi} \rightarrow \overline{f}$, and the phases of the tree and one-loop matrix elements differ by the sum of two constants, one of which flips under CP and one which does not. That is,
\begin{equation}
|\mathcal{M}_{\psi \rightarrow f}|^2 = |\mathcal{M}_{\psi \rightarrow f,t} + \mathcal{M}_{\psi \rightarrow f,1}|^2 = A \left|1 + B e^{i\alpha} e^{i\delta}\right|^2
\end{equation}
where $A = |\mathcal{M}_{\psi \rightarrow f,t}|^2$, $B = \left|\frac{\mathcal{M}_{\psi \rightarrow f,t}}{\mathcal{M}_{\psi \rightarrow f,1}}\right|$, $\alpha$ is CP-odd (for example, the phase coming from the product of coupling constants) and $\delta$ is CP-even. The amplitude of the process $\overline{\psi} \rightarrow \overline{f}$ is just the CP-reversal of this, so
\begin{equation}
|\mathcal{M}_{\overline{\psi} \rightarrow \overline{f}}|^2 = |\mathcal{M}_{\overline{\psi} \rightarrow \overline{f},t} + \mathcal{M}_{\overline{\psi} \rightarrow \overline{f},1}|^2 = A \left|1 + B e^{-i\alpha} e^{i\delta}\right|^2
\end{equation}
and
\begin{equation}
\begin{split}
|\mathcal{M}_{\psi \rightarrow f}|^2 - |\mathcal{M}_{\overline{\psi} \rightarrow \overline{f}}|^2 & = AB \left( e^{i(\alpha + \delta)} + e^{-i(\alpha + \delta)} - e^{i(\alpha-\delta)} - e^{i(-\alpha + \delta)}\right) \\
& = -4 AB \sin(\alpha) \sin(\delta)
\end{split}
\end{equation}
In the examples we study, the CP-even phase comes from the imaginary part of the 1-loop diagrams---that is, the part of the loop integrals where internal particles go on-shell as per the Cutkosky rules.

After some calculation, we find for the process $N_\alpha \rightarrow u^c X$ (Figure \ref{fig:baryoninterference}) the formula \cite{allahverdi2}
\begin{equation}
\epsilon_B =\sum_{i,j,\beta} \frac{  \text{ Im} \left(\lambda_{i\alpha} \lambda^*_{i\beta} \lambda^*_{j\beta} \lambda_{j\alpha} \right)}{24\pi \sum_k \lambda^*_{k\alpha} \lambda_{k\alpha}} \left[3 \mathcal{F}_S \left(\frac{M_\beta^2}{M_\alpha^2}\right) + \mathcal{F}_V \left(\frac{M_\beta^2}{M_\alpha^2}\right) \right].
\end{equation}
Similarly, for the DDDM processes of Figure \ref{fig:dddminterference}, we obtain the same result for $\phi_i \to \psi n_d$ up to the absence of color factors, 
\begin{equation}
\epsilon_d = \sum_{j} \frac{\text{Im} \left[\lambda_i \lambda_j^* \lambda_j^* \lambda_i\right]}{8\pi \sum_k \lambda_k \lambda_k^*} \left[\mathcal{F}_S \left(\frac{M_j^2}{M_i^2}\right) + \mathcal{F}_V \left(\frac{M_j^2}{M_i^2}\right)\right].
\end{equation}
Here $\mathcal{F}_S$ and $\mathcal{F}_V$ are loop functions from self-energy and vertex diagrams respectively:
\begin{equation}
\mathcal{F}_S(x) = \frac{2}{1-x}, \quad \mathcal{F}_V(x) = -2 \left[1-x \text{ ln}\left(1+\frac{1}{x}\right) \right].
\end{equation}
These deviate from expressions in \cite{allahverdi2} because we have included only the scalar decays to fermions, rather than the full set of decays involving all fields in the supermultiplet (i.e.~we assume that SUSY breaking makes scalars heavy enough that no cut scalar lines contribute).

\bibliography{ref}
\bibliographystyle{utphys}

\end{document}